\documentclass[conference]{IEEEtran}
\IEEEoverridecommandlockouts
\usepackage{cite}
\usepackage{amsmath,amssymb,amsfonts}
\usepackage{algorithmic}
\usepackage{graphicx}
\usepackage{textcomp}
\usepackage{tikz}
\usepackage{textcomp}
\usepackage{hyperref}
\usepackage{lipsum}
\usepackage{xcolor}
\def\BibTeX{{\rm B\kern-.05em{\sc i\kern-.025em b}\kern-.08em
    T\kern-.1667em\lower.7ex\hbox{E}\kern-.125emX}}

\newcommand\copyrighttext{%
  \footnotesize \textcopyright 2022 IEEE. Personal use of this material is permitted.
  Permission from IEEE must be obtained for all other uses, in any current or future
  media, including reprinting/republishing this material for advertising or promotional
  purposes, creating new collective works, for resale or redistribution to servers or
  lists, or reuse of any copyrighted component of this work in other works.
}
\newcommand\copyrightnotice{%
\begin{tikzpicture}[remember picture,overlay]
\node[anchor=south,yshift=20pt] at (current page.south) {\fbox{\parbox{\dimexpr\textwidth-\fboxsep-\fboxrule\relax}{\copyrighttext}}};
\end{tikzpicture}%
}

\begin{document}

\title{Methods and Tools for Monitoring Driver’s Behavior\\
}
\author{\IEEEauthorblockN{Muhammad Tanveer Jan}
\IEEEauthorblockA{\textit{College of Engg and Computer Science} \\
\textit{Florida Atlantic University}\\
Boca Raton, USA \\
mjan2021@fau.edu}
\and
\IEEEauthorblockN{Sonia Moshfeghi}
\IEEEauthorblockA{\textit{College of Engg and Computer Science} \\
\textit{Florida Atlantic University}\\
Boca Raton, USA \\
smoshfeghi@fau.edu}
\and
\IEEEauthorblockN{Joshua William Conniff}
\IEEEauthorblockA{\textit{Charles E. Schmidt College of Science } \\
\textit{Florida Atlantic University}\\
Boca Raton, USA \\
fau.jconniff@health.fau.edu}
\and
\IEEEauthorblockN{Jinwoo Jang}
\IEEEauthorblockA{\textit{College of Engg and Computer Science} \\
\textit{Florida Atlantic University}\\
Boca Raton, USA \\
jangj@fau.edu}
\and
\IEEEauthorblockN{Kwangsoo Yang}
\IEEEauthorblockA{\textit{College of Engg and Computer Science} \\
\textit{Florida Atlantic University}\\
Boca Raton, USA \\
yangk@fau.edu}
\and
\IEEEauthorblockN{Jiannan Zhai}
\IEEEauthorblockA{\textit{College of Engg and Computer Science} \\
\textit{Florida Atlantic University}\\
Boca Raton, USA \\
jzhai@fau.edu}
\and
\IEEEauthorblockN{Monica Rosselli}
\IEEEauthorblockA{\textit{Charles E. Schmidt College of Science } \\
\textit{Florida Atlantic University}\\
Boca Raton, USA \\
mrossell@fau.edu}

\and
\IEEEauthorblockN{David Newman}
\IEEEauthorblockA{\textit{Christine E. Lynn College of Nursing} \\
\textit{Florida Atlantic University}\\
Boca Raton, USA \\
dnewma@fau.edu}
\and
\IEEEauthorblockN{Ruth Tappen}
\IEEEauthorblockA{\textit{Christine E. Lynn College of Nursing} \\
\textit{Florida Atlantic University}\\
Boca Raton, USA \\
rtappen@fau.edu}
\and
\IEEEauthorblockN{Borko Furht}
\IEEEauthorblockA{\textit{College of Engg and Computer Science} \\
\textit{Florida Atlantic University}\\
Boca Raton, USA \\
bfurht@fau.edu}
}

\maketitle
\copyrightnotice

\begin{abstract}
In-vehicle sensing technology has gained tremendous attention due to its ability to support major technological developments, such as connected vehicles and self-driving cars. In-vehicle sensing data are invaluable and important data sources for traffic management systems. In this paper we propose an innovative architecture of unobtrusive in-vehicle sensors and present methods and tools that are used to measure the behavior of drivers. The proposed architecture including methods and tools are used in our NIH project to monitor and identify older drivers with early dementia
\end{abstract}


\section{Introduction}
About one in ten older adults in the U.S. have Alzheimer’s disease (AD) and another 15 to 20\% have mild cognitive impairment (MCI), a third of whom will develop dementia within 5 years. Individuals with dementia eventually become unable to perform complex everyday activities including driving so most current driving research focuses on MCI or early stage dementia. Our 5-year project, funded by NIH, focuses on creating an innovative in-vehicle sensors architecture, selecting a large group of older drivers, ages 65 to 85 years old, measuring driver’s behavior during 3-year period, and identify cognitive changes that can lead to early dementia. Leading vehicle manufacturing companies including Volvo \cite{ref1}, Ford
\cite{ref2}	, and Kia \cite{ref3}, have adopted driver alert system applications that alert the drivers to avoid accidents. Most of those systems used the same techniques and algorithms to monitor driver’s behavior but the use of those systems are limited to alerting the driver of any situations and avoid incidents from happening i.e Driver alert systems alert’s the driver if he/she is driving long. Advance Driver assistance systems alerts the driver if vehicle to moving onto other lane or if someone is in their blind spot. Measuring driving behaviors can divided into 3 types \cite{ref12}

\begin{itemize}
  \item Driver Based: Drowsiness, Sensation seeking, impulsive etc.
  \item Driving based: Distraction, attention etc
  \item Qualitative based: Speeding, braking, lane changing etc.
\end{itemize}
Among above mentioned distraction and drowsiness are major behavior that changes with time and can be used to find key insights into driving of participants. One of them is distraction.
Some of the usual distractions during driving are when drivers are looking off road and texting and using phone \cite{ref7,ref8,ref9}. Some of the metrics that are used to measure distraction are head- pose \cite{ref10} and gaze patterns \cite{ref11}

\section{IN-VEHICLE SENSOR ARCHITECTURE }


We designed an innovative architecture of unobtrusive in- vehicle sensors. The concept of utilizing in-vehicle sensors to measure the behavior of drivers and detect cognitive change   is in itself innovative and reflective of the rapid  development of these sensors and their application for monitoring driver behavior. Published reports of the proposed sensors are limited, particularly in regard to small sample size, duration  of the testing and the number of  sensors  and/or  cognitive tests utilized. Furthermore, few prospective studies examined patients with abnormal aging (MCI and early dementia) and there have been no studies including a culturally diverse sample. Our proposal addresses these limitations with a fully powered study, complete package of unobtrusive sensors and array of cognitive tests including the LASSI-L and others that are particularly sensitive to the early changes in cognition. Furthermore, the algorithms to measure Driving Behavior In- dices are unique and will contribute significantly to application of the results in real-world settings. The specific technical innovations of this project include the following
\begin{itemize}
    \item an adaptive driver behavior data sampling technique to improve the efficiency of the data collection for driver behavior analytics,
    \item the data fusion of telematics, vision-sensing, and the environmental data to define highly detailed driver behavior indices,
    \item the creation of a scalable spatial network query processing platform for heterogeneous data sources.
\end{itemize}
\par
\begin{figure}[htbp]
\centerline{\includegraphics[width=8cm]{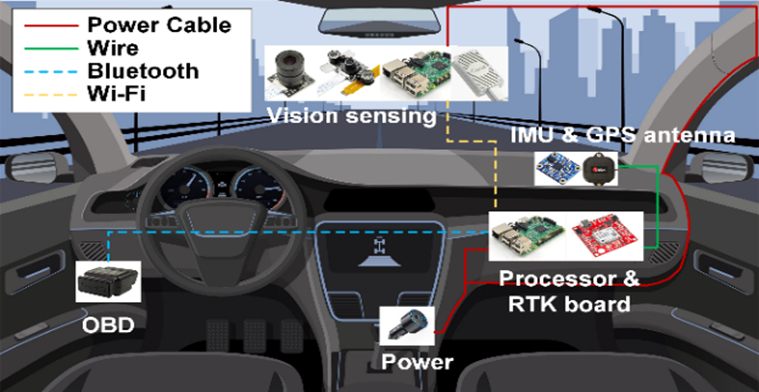}}
\caption{Architecture of Vision Sensor.}
\label{p1}
\end{figure}

The architecture of the proposed in-vehicle sensors is shown in fig \ref{p1}. Each in-vehicle sensor is comprised of two distributed sensing units: 
\begin{itemize}
    \item in-vehicle vision sensing unit,
    \item in-vehicle telemetry unit.
\end{itemize}
For example, the GPS clock information will share data throughout the in-vehicle sensor networks to achieve precise time synchronization. The proposed sensors supports powerfull computing resources for real-time, customized onboard data processing, data-fusion, and machine-vision. Customized sensor enclosures are designed for the sensors to support a cooling system and to minimize the size of sensor hardware. Each in-vehicle sensor has local data storage to store in- vehicle sensing data. The data are uploaded to a central database during participants’ quarterly cognitive testing visits. In-Vehicle sensing units are the main components that are used in this study, there are two types of sensing units that are used.

\subsection{Vision Sensors}
The major components of the vision sensing unit are forward-facing camera, driver-facing camera, and MDVR device to store data. The vision sensor unit is mounted on the windshield, The vision sensors use computer vision algorithms to unobtrusively
\begin{itemize}
    \item track eye and head movements,
    \item read facial micro-expression,
    \item driving situation awareness (e.g., traffic light classification, taillights of front vehicles, and lane marking, stop sign, and vehicle detection.
\end{itemize}
Table \ref{des1} shows indices measured by driver-facing and front  cameras. We describe next several methods and tools that are using vision sensors to measure driver’s behavior.

\begin{table}[htbp]
\caption{Driver behavior indices measure by two in-vehicle cameras}
\begin{center}
\begin{tabular}{p{0.2\linewidth} | p{0.2\linewidth} | p{0.5\linewidth}}
\hline
\textbf{Camera} & \textbf{Feature} & \textbf{description} \\
\hline
\textbf{Driver's Camera} & Face Detection & The AI algorithm detects the face of the driver and the driver’s facial features. \\

 & Eyes Detection & The AI algorithm detects the eyes and analyzed weather driver eyes are open or closed.\\

 & Yawning & Yawning is based on eyes and mouth detection.\\

 & Distraction & Distraction is based on the head-pose angle. Head-pose estimation technique was applied but the initial angle is adjusted due to the placement of the camera on dashboard\\

 & Smoking\&Phone Use & Both of these indices use AI object detection algorithms to detection. \\
\hline
\textbf{Front Camera} & Traffic Sign Detection & The AI algorithm detects the traffic signals and monitor if the driver passes on red sign or not \\

& Object Detection & The AI algorithm detects objects on the roads,\par such as pedestrian or cyclist crossing the road, curbs or barriers, nearby vehicles, and others. \\

 & Lane Crossing & AI algorithm that can detect lane departure  \\

 & Near-Collision & AI algorithm that can detect object or vehicle at a certain distance from it.  \\

 & Pedestrian detection & AI algorithm that can detect whether the driver yield at for the passenger crossing.  \\
\hline
\end{tabular}
\label{des1}
\end{center}
\end{table}

\subsubsection{Region of Interest and Face Detection: }
\begin{figure}[htbp]
\centerline{\includegraphics[width=8cm]{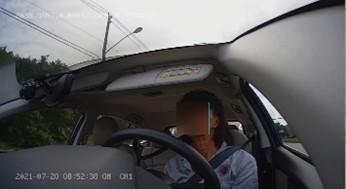}}
\caption{Face Detection based on region-of-interest.}
\label{p2}
\end{figure}

\begin{figure}[htbp]
\centerline{\includegraphics[width=8cm]{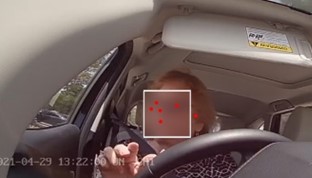}}
\caption{6 Facial landmark detection.}
\label{p3}
\end{figure}

Multiple techniques are used for face detection based on the region of interest. Region-of-Interest is identified by considering different metrics such are angle of the camera, height of the driver, place of the camera. Number of drivers are large in number so in order to detect the RoI and Face we used them interchangeably such that we use a less accurate but fast face detection library to have a rough estimate of the location of the face and then based on that location. RoI is selected with maximum area for the face or any false positive that it may result in and also to rule out the passenger face. After RoI is identified, it is passed to more accurate face detector to have an exact location with facial landmarks. Fig \ref{p2} \& Fig \ref{p3} shows is results from our vision sensor system

\subsubsection{Eyes and Mouth Detection: }
\begin{figure}[htbp]
\centerline{\includegraphics[width=8cm]{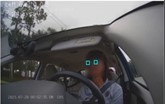} }
\caption{Eyes Detection.}
\label{p4}
\end{figure}

\begin{figure}[htbp]
\centerline{\includegraphics[width=8cm]{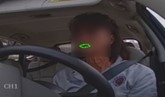}}
\caption{Mouth Detection.}
\label{p5}
\end{figure}

Eyes and mouth detection are two more features implemented in our vision system. Both of these features are detected separately and are based on HAAR-Cascade classifiers. The functions return a bounding box for the detected area of eyes and mouth and can be passed as an array or tuple for RGB or grayscale image for further  use. Apart from that we also implemented emotion recognition but were excluded due not sufficient accuracy\cite{ref14}.  Fig \ref{p4} \& fig \ref{p5} illustrates mouth and eyes detection in our video sensor system.

 \subsubsection{Lane Detection}
 Lane detection is another feature of the video sensor system that uses deep learning model to detect lane while driving. It takes image or video as an input and display a green marker on the current lane on which the driver is driving. Rain detection is also included in this framework that used machine learning model to predict the rain

\subsection{Telemactic Sensors}
The major sensing components of the telematics unit are 
\begin{itemize}
    \item an on-chip RTK GNSS module,
    \item an Inertial Measurement Unit (IMU),
    \item an on-board diagnostics (OBD)
\end{itemize}
They enable obtaining high-precision positioning data, real-time vehicle state information (e.g., engine RPM, vehicle speed, and pedal position), and vehicles’ dynamic motions. 

\subsubsection{High-precision Positioning:}
 An on-chip RTK GNSS module allows capturing a complete picture of vehicles’ high- precision positioning data, making in-depth driver behavior analytics at lane and intersection levels possible. This new technology provides highly precise positioning data compared to typical GPS mod- ules (accuracy about 4.9 m). The precise vehicle localization achieved by this new technology is important to capture lane deviations, travel patterns in parking lot areas, and detailed turning behaviors.
\subsubsection{Onboard Diagnostics:} A Bluetooth OBD scanner will obtain real-time data streams from the vehicle’s Controller Area Network (CAN bus). Standardized OBD-II protocol facilitates easy data translation into human-readable formats based on standardized OBD-II PIDs without reverse engineering work.
\subsubsection{Motion and Orientation:} IMU sensors, consisting of tri-axial accelerometers, gyroscopes, and magnetometers, are used to capture vehicles’ dynamic motions and orientations. They estimate the harsh acceleration, braking, and cornering of vehicles, which are one of the popular indicators for driver behaviors. IMU sensors also identify driving over potholes and raised pavement markers, allowing for analysis of how drivers react to unexpected pavement defects and lane departure in conjunction with the other telematics data.

\section{Driver Behavior Indices}
Our objective is to measure and monitor changes of Driver Behavior Indices (DBIs) using the vision and telematics sensor data. The selection of the DBIs is designed to reflect older drivers’ cognitive function and driving performance. The DBIs is evaluated for each driver and summarized on a daily, weekly, and monthly basis. DBIs are classified into four categories. Examples of DBIs are shown in Table \ref{des2}. 

\begin{table}[htbp]
\caption{Driver behavior indices}
\begin{center}
\begin{tabular}{p{0.2\linewidth} | p{0.3\linewidth} | p{0.4\linewidth}}
\hline
\textbf{Categories} & \textbf{Driver Behavior Indices} & \textbf{Data Analytics} \\
\hline
\textbf{Travel Pattern} 
& number of trips, driven miles, miles on the highway, miles during the night, daytime, and severe weather, miles on highway, etc.
& map-matching,  data queries, map data, weather data
\\

\textbf{Abnormal Driving}
& wayfinding, getting lost, ignore traffic signals and signs, near-collision events, distraction, drowsiness, etc.
& machine vision, shortest path, outlier detection, trajectory clustering, frequent graph mining
\\

\textbf{Reaction Time}
& reaction time to traffic light change, front-vehicle taillight, pothole, etc.
& vision sensing, data-fusion, vibration analysis, machine learning
\\

\textbf{Braking Patterns}
& eye movements and IMU data due to stop signs, traffic signal, trail light, losing focus, potholes, etc.
& signal processing, gaze estimation, data mining, machine vision
\\
\hline
\end{tabular}
\label{des2}
\end{center}
\end{table}
\begin{figure}[htbp]
\centerline{\scalebox{0.40}{\includegraphics{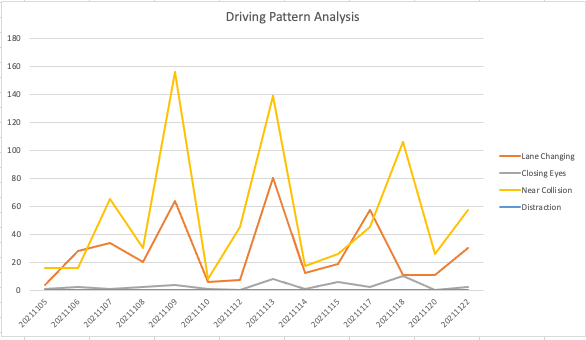}}}
\caption{Driving Pattern driver \# 1}
\label{p6}
\end{figure}

\begin{figure}[htbp]
\centerline{\scalebox{0.40}{\includegraphics{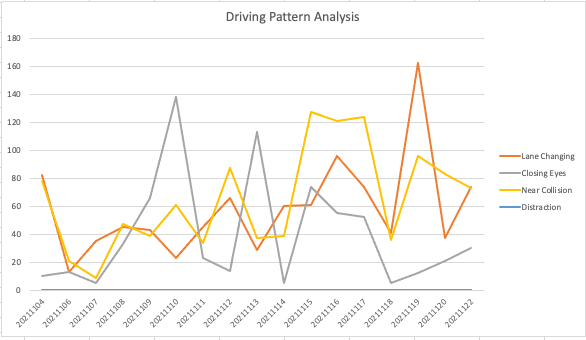}}}
\caption{Driving Pattern driver \# 2}
\label{p7}
\end{figure}

For illustration, fig \ref{p6} and \ref{p7} shows the driving pattern of a  two senior drivers for 2-weeks period based on the data obtained from in-vehicle cameras that measured the number of times the driver closed eyes, number of distractions, crossing lines, and near collision events.details of the the data can be reviewed in previous works that are published is same domain \cite{ref13}

\section{Conclusion}
We presented an innovative architecture of an in-vehicle sensor system consisting of vision and telemetry sensors and includes a set of AI algorithms to measure driver behavior indices. The system was already installed in about 70 cars driven by older drivers in Florida with the objective to monitor and detect cognitive changes in these drivers. The final goal  of the project is to identify those drivers that their cognitive changes imply early dementia. The detailed results of the study will be published soon.

\section{Acknowledgment}
The project mentioned in section I is supported by Award Number GT003128-NIH from the National Institute of Health. Its contents are solely the responsibility of the authors and do not necessarily represent the official views of the National Institutes of Health.


\end{document}